\documentclass[aps,prb,twocolumn,preprintnumbers]{revtex4}
\usepackage{graphicx}
\usepackage{amstext}
\usepackage{array}
\usepackage{amssymb}
\usepackage{amsmath}
\usepackage{color}
\newcommand{\be}{\begin{equation}}
\newcommand{\ee}{\end{equation}}
\newcommand{\bea}{\begin{eqnarray}}
\newcommand{\eea}{\end{eqnarray}}
\newcommand{\Eq}[1]{Eq.\,(\ref{#1})}
\newcommand{\Fig}[1]{Fig.\,\ref{#1}}
\newcommand{\Sec}[1]{Sec.\,\ref{#1}}
\newcommand{\Tab}[1]{Table\,\ref{#1}}
\newcommand{\Cite}[1]{~[\onlinecite{#1}]} 
\newcommand{\Onlinecite}[1]{Ref.\,[\onlinecite{#1}]} 

\newcommand{\br}{\mathbf{r}}

\newcommand{\heps}{\hat{\boldsymbol{\varepsilon}}}
\newcommand{\epsD}{\varepsilon_{\rm D}}
\newcommand{\hsigma}{\hat{\boldsymbol{\sigma}}}
\newcommand{\hmu}{\hat{\boldsymbol{\mu}}}

\newcommand{\mepsi}{\varepsilon_j''}
\newcommand{\mepsr}{\varepsilon_j'}

\newcommand{\mepst}{\Delta\varepsilon_j}
\newcommand{\mepsrt}{\Delta\varepsilon_j'}
\newcommand{\mepsit}{\Delta\varepsilon_j''}

\newcommand{\eps}{\varepsilon}
\newcommand{\epsi}{\varepsilon''}
\newcommand{\epsr}{\varepsilon'}

\newcommand{\epst}{\Delta\varepsilon}
\newcommand{\epsrt}{\Delta\varepsilon'}
\newcommand{\epsit}{\Delta\varepsilon''}

\newcommand{\feps}{\varepsilon(\omega_j)}
\newcommand{\fepsi}{\varepsilon''(\omega_j)}
\newcommand{\fepsr}{\varepsilon'(\omega_j)}
\newcommand{\func}[1]{g_{#1}(\omega_j)}
\newcommand{\funci}[1]{g_{#1}''(\omega_j)}
\newcommand{\funcr}[1]{g_{#1}'(\omega_j)}
\newcommand{\sigmar}[1]{\sigma_{#1}'}
\newcommand{\sigmai}[1]{\sigma_{#1}''}

\newcommand{\pade}[2]{\frac{\partial #1}{\partial #2}}

\begin{document}
\title{Optimizing the Drude-Lorentz model for material permittivity: \\Method, program, and examples for gold, silver, and copper}
\author{H.\,S. Sehmi}
\author{W. Langbein}
\author{E.\,A. Muljarov}
\affiliation{School of Physics and Astronomy, Cardiff University, Cardiff CF24 3AA,
United Kingdom}
\begin{abstract}
Approximating the frequency dispersion of the permittivity of materials with simple analytical functions is of fundamental importance for understanding and modeling the optical response of materials and resulting structures.
In the generalized Drude-Lorentz model, the permittivity is described in the complex frequency plane by a number of simple poles having complex weights, which is a physically relevant and mathematically simple approach: By construction, it respects causality represents physical resonances of the material, and can be implemented easily in numerical simulations. We report here an efficient method of optimizing the fit of measured data with the Drude-Lorentz model having an arbitrary number of poles. We show examples of such optimizations for gold, silver, and copper, for different frequency ranges and up to four pairs of Lorentz poles taken into account. We also  provide a program implementing the method for general use.
\end{abstract}
%
%
\date{\today}
\maketitle

\section{Introduction}

The interest in nanotechnology  has increased greatly over the past decade, particularly in nanophotonics, which exploits optical properties of structures on the nanoscale, composed of different materials. In order to design photonic structures and predict and optimize their properties, such as optical field enhancement, chirality, or enhanced radiative emission via the Purcell effect, the electromagnetic  response of the underlying materials has to be simulated. An effective medium approach of the optical response is suited for many structures in which nonlocal effects can be neglected. The properties describing the linear optical response of non-chiral media are the frequency dependent permittivity tensor $\heps(\omega)$ and permeability tensor $\hmu(\omega)$. In most relevant natural materials, the permeability is close to unity, so that we concentrate here on $\heps(\omega)$. However, the method is applicable equivalently to $\hmu(\omega)$ or, in general, to other material response functions.

Using an analytical model of $\heps(\omega)$, which contains only simple poles, is motivated by physical arguments, such as the presence of resonances in the material self energy and response functions. Furthermore, this form of the permittivity can be efficiently implemented in numerical methods, such as the finite difference in time domain (FDTD) method \Cite{VialJPD07}, and in the more analytic and rigorous approaches, such as the dispersive resonant-state expansion\Cite{MuljarovPRB16}. The pole structure of the permittivity naturally includes a zero-frequency pole of the Ohm's law dispersion, which however works well only in the long wavelength limit and is not suited to describe the material properties in the optical range. Real metals are much better described by the Drude model\Cite{JohnsonPRB72}, which takes into account the finite mass of the charge carriers. Adding real-valued Lorentz components\Cite{RakicAO98} to the Drude model is suited to represent electronic interband transitions. A further refinement of the model uses complex weights (residues) of the Lorentz poles\Cite{EtchegoinJCP06,EtchegoinJCP07}. This is known in some of the literature as the critical point model\Cite{VialJPD07, VialAPB08,VialAPA11}. We use this generalization in the present work and call it a Drude-Lorentz (DL) model.

A fit of the material permittivity with the DL model has been performed in a number of publications
\Cite{EtchegoinJCP06,EtchegoinJCP07,VialJPD07, VialAPB08,VialAPA11}
for its further use in  FDTD solvers. However, the experimental errors available in the literature\Cite{JohnsonPRB72} have not been taken into account in those fits. In the present paper, we provide an efficient algorithm of fitting  experimental data, using available errors, with the DL model with an arbitrary number of Lorentz poles. This algorithm combines an exact analytical approach for determining the linear parameters of the model, with a numerical solver for optimizing its nonlinear parameters. We illustrate the resulting pole positions and their weights  in the complex plane to give some physical insight how the model approximates the electronic transitions in real materials.

The paper is organized as follows. Section~\ref{DL}  introduces a generalized DL model of the permittivity. Section~\ref{sec:ExpData} relates the experimental errors of the measured refractive and absorption indices to the errors of the complex permittivity used for the fit. The fit procedure, including the analytical and numerical optimization of parameters of the DL model and the algorithm for determining appropriate starting values in the gradient descent minimization procedure is described in Sec.\ref{sec:Opt}. Results of the fit are provided in \Sec{sec:Results} for gold and in the Appendix for silver, copper, and monocrystalline gold.

\section{Drude-Lorentz model}
\label{DL}

Quite generally, the permittivity $\heps(\br,\omega)$ can be treated as an analytic function in the complex frequency plane, having a countable number of simple poles and therefore, according to the Mittag-Leffler theorem, can be expressed as
\be
\heps(\omega)=\heps_\infty+\sum_j\frac{i\hsigma_j}{\omega-\Omega_j}\,,
\label{eqn:eps}
\ee
where $\heps_\infty$ is the high-frequency value of the permittivity and $\Omega_j$ are the  resonance frequencies, which are the poles of the permittivity, determining its dispersion, with the weight tensors $\hsigma_j$ corresponding to generalized conductivities of the medium at these resonances. The Lorentz reciprocity theorem requires that all tensors in \Eq{eqn:eps} are symmetric, and the causality principle requires that $\heps(\omega)$ has no poles in the upper half plane of $\omega$ and that $\heps^\ast(\omega)=\heps(-\omega^\ast)$\Cite{LandauLifshitzV8Book84}. Therefore, for a physically relevant dispersion, each pole of the permittivity with a positive real part of $\Omega_j$ has a partner at $\Omega_{-j}=-\Omega^\ast_j$ with $\hsigma_{-j}=\hsigma^\ast_j$. Poles with zero real part of $\Omega_j$ have real $\hsigma_j$. For simplicity, we assume in the following an isotropic response, such that the conductivities and thus the permittivity are described by scalars. We note however that it is straightforward to extend the presented treatment to a nonisotropic response.

We first separate the poles with zero real part of the frequency, which describe the conductivity of materials in the Drude model:
\be
\epsD(\omega) = \varepsilon_\infty + \frac{i\sigma}{\omega} - \frac{i\sigma}{\omega+i\gamma} = \varepsilon_\infty - \frac{\gamma\sigma}{\omega(\omega+ i\gamma)}\,,
\label{Drude}
\ee
where $\varepsilon_\infty$ is the permittivity at high frequencies and $\sigma$ is the real DC conductivity. The pole at zero frequency represents Ohm's law, corresponding to the $\omega^{-1}$ low-frequency limit of the dispersion. Together with the second pole, at $-i\gamma$, it provides the $\omega^{-2}$ high-frequency asymptotics, originating from the nonzero mass of the charge carriers. In real materials, the carrier mass and the damping can show a frequency dependence, which is not included in the Drude model. To describe such effects, the DC conductivity can be split \Cite{AllenPRB77} into several Drude contributions, with fractions $\eta_d$ and dampings $\gamma_d$, so that
\be
\epsD(\omega) = \varepsilon_\infty +\frac{i\sigma}{\omega} - \sigma \sum_{d=1}^{D}\frac{i\eta_d}{\omega+i\gamma_d}\,,
\label{DrudeEx}
\ee
where $\sum_{d=1}^{D} \eta_d=1$. Adding the poles $\Omega_k$ with nonzero real part, which are called Lorentz poles and describe material resonances at finite resonance frequencies, such as phonons or electronic interband transitions, we arrive at
\be
\varepsilon(\omega)=\epsD(\omega) +\sum\limits_{k=1}^L\left(\frac{i\sigma_k}{\omega-\Omega_k} + \frac{i\sigma_k^*}{\omega+\Omega_k^*}\right)\,,
\label{eqn:eps2}
\ee
where $L$ is the number of pairs of Lorentz poles. The generalized conductivities $\sigma_k = \sigmar{k}+i\sigmai{k}$ are complex. We denote real and imaginary parts of complex quantities with prime and double prime, respectively, and keep using this notation throughout the paper.

The model of the permittivity $\eps(\omega)$ given by the analytic function \Eq{eqn:eps2} with $\Omega''_k\leq 0$ respects the constrain of causality by construction. The parameters of the model, which are the conductivities and the resonance frequencies, have to be determined from the experimentally measured data.

\section{Experimental Data and Errors}
\label{sec:ExpData}
In typical experiments\Cite{JohnsonPRB72}, the refractive index $n(\omega)$ and absorption index $\kappa(\omega)$ are determined at a number of real frequencies $\omega_j$ providing $n_j=n(\omega_j)$ and  $\kappa_j=\kappa(\omega_j)$. The measured values are assumed here to have an error defined by the root-mean square (RMS) deviation, $\Delta n_j$ and $\Delta \kappa_j$, respectively.
We will see in \Sec{sec:Opt} that it is computationally advantageous to find the model parameters by minimizing the deviation of $\varepsilon$, and not of $n$ and $\kappa$. We therefore calculate $\varepsilon=(n+i\kappa)^2$ and determine its RMS error $\Delta\varepsilon$ by assuming statistically independent errors $\Delta n$ and $\Delta \kappa$, which yields
\begin{align}
	\epsrt  &= \sqrt{\left(\pade{\epsr}{n} \Delta n \right)^2 + \left( \pade{\epsr}{\kappa}\Delta \kappa\right)^2}\nonumber\\
                      &= 2\sqrt{(n\Delta n)^2+(\kappa \Delta \kappa)^2}\,,
\label{RMS1}
                      \\
	\epsit  &= \sqrt{\left(\pade{\epsi}{n} \Delta n \right)^2 + \left( \pade{\epsi}{\kappa}\Delta \kappa\right)^2}\nonumber\\
                      &= 2\sqrt{(\kappa \Delta n)^2+(n\Delta \kappa)^2}\,.
	\label{RMS2}
\end{align}
We then define $\mepst=\epst(\omega_j)$, treating all quantities in Eqs.\,(\ref{RMS1}) and (\ref{RMS2}) as functions of $\omega$. We assume that the frequencies $\omega_j$ are sorted in ascending order, and that the minimum (maximum) frequency is $\omega_1$ ($\omega_N$).

\section{Optimization}
\label{sec:Opt}

With the analytic model \Eq{eqn:eps2} of the permittivity, the task of fitting the experimental data reduces to finding the parameters of the model which minimize the error weighted deviation $E$ between the analytic and the measured values of $\eps$, as this maximizes the probability of the model given the data. Assuming Gaussian errors, we use the squared deviation, weighted with the RMS errors:
\be
E = \sum\limits_{j=1}^{N}  \left( \frac{\fepsr - \mepsr}{\mepsrt} \right)^2 + \left( \frac{\fepsi - \mepsi}{\mepsit} \right)^2\,,
\label{eqn:Error}
\ee
where $\eps_j$ are experimental values and $\Delta\eps_j$ are the corresponding errors.
Considering that typical experimental data sets consist of tens to hundreds of points, and $\eps(\omega)$ is an analytic function of $\omega$ with a large number of parameters, typically in the order of ten, a robust and efficient algorithm is needed. To achieve this goal, we first make use of an exact, analytical minimization with respect to the parameters in which $\eps$ is linear -- these are all the conductivities and $\eps_\infty$. This is the reason why it is advantageous to fit $\varepsilon$ instead of the complex refractive index $n+i\kappa$, as for the latter none of the parameters is linear. Then for the rest of the parameters, in which $\eps$ is nonlinear -- these are the pole frequencies -- we use an iterative minimization with a gradient decent and a suited selection of starting points.

\subsection{Exact minimization over linear parameters}
\label{sec:exact}

An exact minimization of the RMS deviation is available for all the parameters in which the model is linear. To make this linear dependence more clear, we write the permittivity as
\be
\feps=\sum_{l=0}^{2L+D} A_l \func{l}
\label{eps3}
\ee
with $1+D+2L$ {\it real linear} parameters $A_l$ and the related complex functions $\func{l}$ as given in \Tab{tab:epsfunc}.

\begin{table}[h]
  \begin{tabular}{c | >{$}c<{$} | >{$}c<{$}}
  $l$  & A_l & g_l(\omega)\\
    \hline
\rule{0pt}{3ex}
 0 & \varepsilon_\infty  & 1 \\
\rule{0pt}{3ex}
    $d$ & \sigma\eta_d  &\displaystyle -\frac{\gamma_d}{\omega(\omega+ i\gamma_d)}\\
\rule{0pt}{4ex}
    $2k+D-1$   & \sigmar{k}   &\displaystyle \frac{i}{\omega-\Omega_k} + \frac{i}{\omega+\Omega_k^*}\\
\rule{0pt}{4ex}
   $ 2k+D$     & \sigmai{k}              & \displaystyle\frac{-1}{\omega-\Omega_k} + \frac{1}{\omega+\Omega_k^*} \\
  \end{tabular}
  \caption{Linear parameters $A_l$ and related functions $\func{l}$ used in the model, with the integers $d=1..D$ and $k=1..L$.}
  \label{tab:epsfunc}
\end{table}

Minimization of the total error $E$, given by \Eq{eqn:Error}, with respect to $A_l$ can be done analytically by setting the derivatives to zero,
\begin{multline}
\pade{E}{A_l} = 2\sum\limits_{j=1}^N \left( \frac{\funcr{l}\left(\sum\limits_m A_m \funcr{m} - \mepsr \right)}{(\mepsrt)^2} \right.\\
                       + \left. \frac{\funci{l}\left(\sum\limits_m A_m \funci{m} - \mepsi \right)}{(\mepsit)^2} \right) = 0.
\end{multline}

These provide a set of $1+D+2L$ linear equations for $A_m$ which can be written as
\be
\sum_{m=0}^{2L+D} H_{lm} A_m = B_l,
\label{matreq}
\ee
where
\begin{align}
H_{lm} &=  \sum\limits_{j=1}^{N} \left( \frac{\funcr{l}\funcr{m}}{(\mepsrt)^2} + \frac{\funci{l}\funci{m}}{(\mepsit)^2} \right)\,,\\
B_l &=  \sum\limits_{j=1}^{N} \left( \frac{\funcr{l}\mepsr}{(\mepsrt)^2} + \frac{\funci{l}\mepsi}{(\mepsit)^2} \right)\,.
\end{align}
Equation~(\ref{matreq}) can be solved using standard linear algebra packages with a computational complexity of $(1+D+2L)^2$, which is smaller than the complexity of $N(1+D+2L)^2$ for calculating $H_{lm}$ and $B_l$ for typical sizes of datasets and number of poles. We can fix the value of $\varepsilon_\infty$ if necessary, removing it from the set of linear parameters, by subtracting our chosen value $\varepsilon_\infty$ from $\feps$ (see an example in \Tab{tab:para}).

\subsection{Minimization over nonlinear parameters}

Using the values of $A_l$ found in \Sec{sec:exact} by exact  minimization of $E$, we now define, via \Eq{eqn:Error}, a new error function $\widetilde{E}$, which has been already minimized with respect to the linear parameters  $A_l$ and depends only on the nonlinear parameters, which are the Drude dampings $\gamma_d$ and the complex frequencies $\Omega_k$ of the Lorentz poles. Overall, there are $D+2L$ real parameters over which $\widetilde{E}$ has to be minimized.  To represent the average deviation of the model from the measured data points relative to their experimental RMS error, we introduce
\be
S= \sqrt{\frac{\widetilde{E}}{2N}}\,.
\label{S}
\ee
A fit to the experimental data has two sets of independent errors relative to the correct $\varepsilon(\omega)$: the error of the measurements and the errors of the fit. For a fit which is equal to the correct $\varepsilon(\omega)$, we expect, by definition, $S=1$. If instead the magnitude of both errors are the same, and they are uncorrelated, we expect $S=\sqrt{2}$. Therefore, for a fit dominated by the measurement errors, the $S$ values are expected to be close to unity, below $\sqrt{2}$. Furthermore, we note that there are $1+2D+4L$ fitting parameters and $2N$ data values, which can be of comparable number. Therefore, there are only $2N-1-2D-4L$ values which cannot be exactly fitted by the model function. Indeed, the set of the fit conditions is overdetermined and thus provides a finite error of the fit, resulting in finite values of $S$ below unity. Specifically, we would expect for the best fit a value of $S =\sqrt{1-(1+2D+4L)/(2N)}$. When the expression under the square root is zero or negative, it is possible to fit the data exactly, i.e. $S$ can approach zero -- we will see examples of this later.

During the minimization, we found instances (specifically when fixing $\varepsilon_\infty$) where the pole frequency and the corresponding weight diverged simultaneously with fixed ratio, representing a frequency-independent permittivity component $i\hsigma_j/\Omega_j$. Furthermore, we observed poles at nearly equal positions, or Lorentz poles on the imaginary axis. All these situations correspond to local minima of $\widetilde{E}$ which should be avoided. We also found poles with positive imaginary part, which are not compatible with causality of the response. In order to avoid the corresponding un-physical pole properties while not significantly compromising the resulting error $S$, we minimize not $\widetilde{E}$, but $\widetilde{E}\zeta$ instead, where
\begin{align} 
\alpha_i&=\left( 1+ \frac{s_1^2 \delta^2}{\left| \Omega_i' \right|^2}\right)\left( 1+ \frac{s_2^2 \delta^2}{\left| \Omega_i'' \right|^2}\right)\\
\beta_i&=\prod\limits_{j>i}\left( 1+ \frac{s_3^2 \delta^2}{\left| \Omega_i'-\Omega_j'\right|^2 + \left| \Omega_i''-\Omega_j''\right|^2}\right)\\
\zeta&=\prod_i \alpha_i\beta_i\zeta_i,\,\,  \zeta_i=\left\{ \begin{array}{ll} 1 & \mbox{for}\,|\Omega_i|<\omega_N\,, \\  1+s_4^2\left(\frac{|\Omega_i|}{\omega_N}-1\right)^2& \mbox{else}\,. \end{array} \right. \label{eqn:limit}
\end{align}
The Drude poles are included in the product \Eq{eqn:limit} using their pole frequencies $\Omega=-i\gamma_d$. The parameter $\delta$ denotes the maximum width between data points. The factors $s_1$ and $s_2$ determine the strength of the repulsion of the Lorentz poles from the imaginary and real axes respectively, $s_3$ determines the strength of the repulsion between Lorentz poles, and $s_4$ determines the strength of the repulsion for absolute pole position larger than $\omega_N$. We used $s_1=0.2$, $s_2=0.5$, $s_3=0.2$ and $s_4=0.04$ for the results shown in this work. Generally, the repulsion parameters are increased from zero to suppress unphysical pole positions and to find the global minimum of $\widetilde{E}$. 

\begin{figure}[t]
\includegraphics[width=0.9\columnwidth]{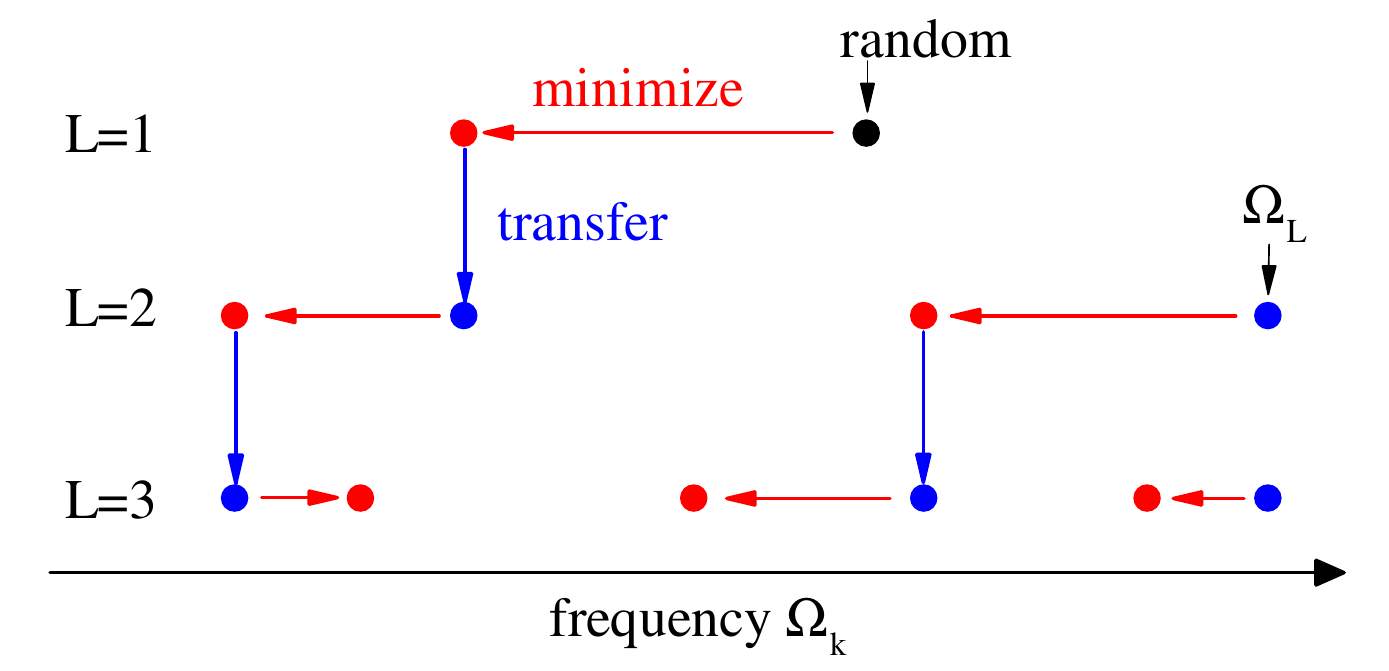}
\caption{Sketch of the procedure for choosing start values for Lorentz pole frequencies $\Omega_k$, with an increasing number of Lorentz pole pairs $L$.}
\label{fig:Guess}
\end{figure}

To minimize $\widetilde{E}\zeta$ over the $D+2L$ nonlinear parameters we use known minimization algorithms based on the gradient decent (implemented in MATLAB as function `fminunc'). The main challenge is to select suited starting points for the parameters, from which the algorithm finds local minima. The starting points should be selected in a way that the global minimum amongst the local minima is found.

The complexity of the problem depends on the number of Drude poles $D$ and Lorentz poles $L$. For $L=0$ and $D=1$, only a single parameter $\gamma_1$ has to be varied, which results in a reliable convergence towards the global minimum independent of the choice of its start value. Increasing $D$ to $D+1$, we use as starting value $\gamma_{D+1}=2\gamma_D$.

For $L=1$, we have an additional pair of Lorentz poles given by a single complex parameter $\Omega_1$.
For the starting value of $\Omega_1$, we use a random logarithmic distribution within the range of the measured data, specifically
\be
\Omega_1 = \omega_1 \left( \frac{\omega_N}{\omega_1} \right)^Y - i (\omega_N - \omega_1)N^{Y'-1}
\ee
where $Y$ and $Y'$ are random numbers with a uniform distribution between 0 and 1. The minimization is repeated with different starting points until at least three resulting $S$ values are equal within 10\%, and the parameters for the lowest $S$ are accepted as global minimum.

\begin{figure}[t]
\includegraphics*[width=\columnwidth]{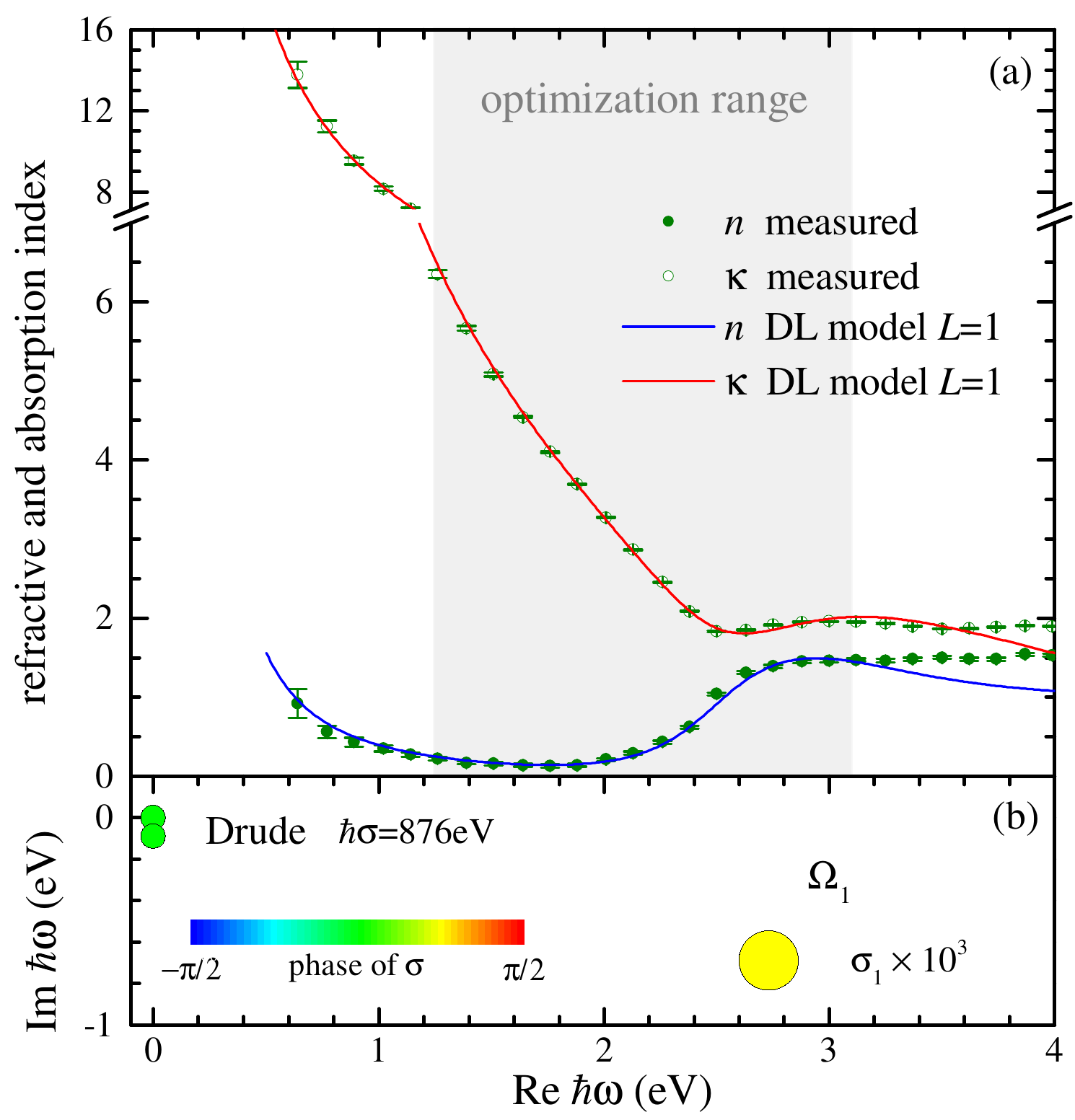}
	\caption{(a) Refractive index $n$ and absorption index $\kappa$ of gold according to\Cite{JohnsonPRB72} (circles and error bars) and the DL model \Eq{eqn:eps2} for $L=1$ (solid lines) as functions of the photon energy $\hbar\omega$. The fit is optimized for the range $1.24\leqslant \hbar\omega \leqslant3.1$\,eV. (b) Pole positions $\Omega_j$ and weights $\sigma_j$ of the fitted $\eps(\omega)$. The circle area is proportional to $|\sigma_j|$, and color gives the phase of $\sigma_j$ as indicated. For the Lorentz poles, $\sigma_j$ is multiplied by a factor of 1000 for clarity.}
	\label{fig:JCAuL1}
\end{figure}

\begin{figure}[t]
\includegraphics*[width=\columnwidth]{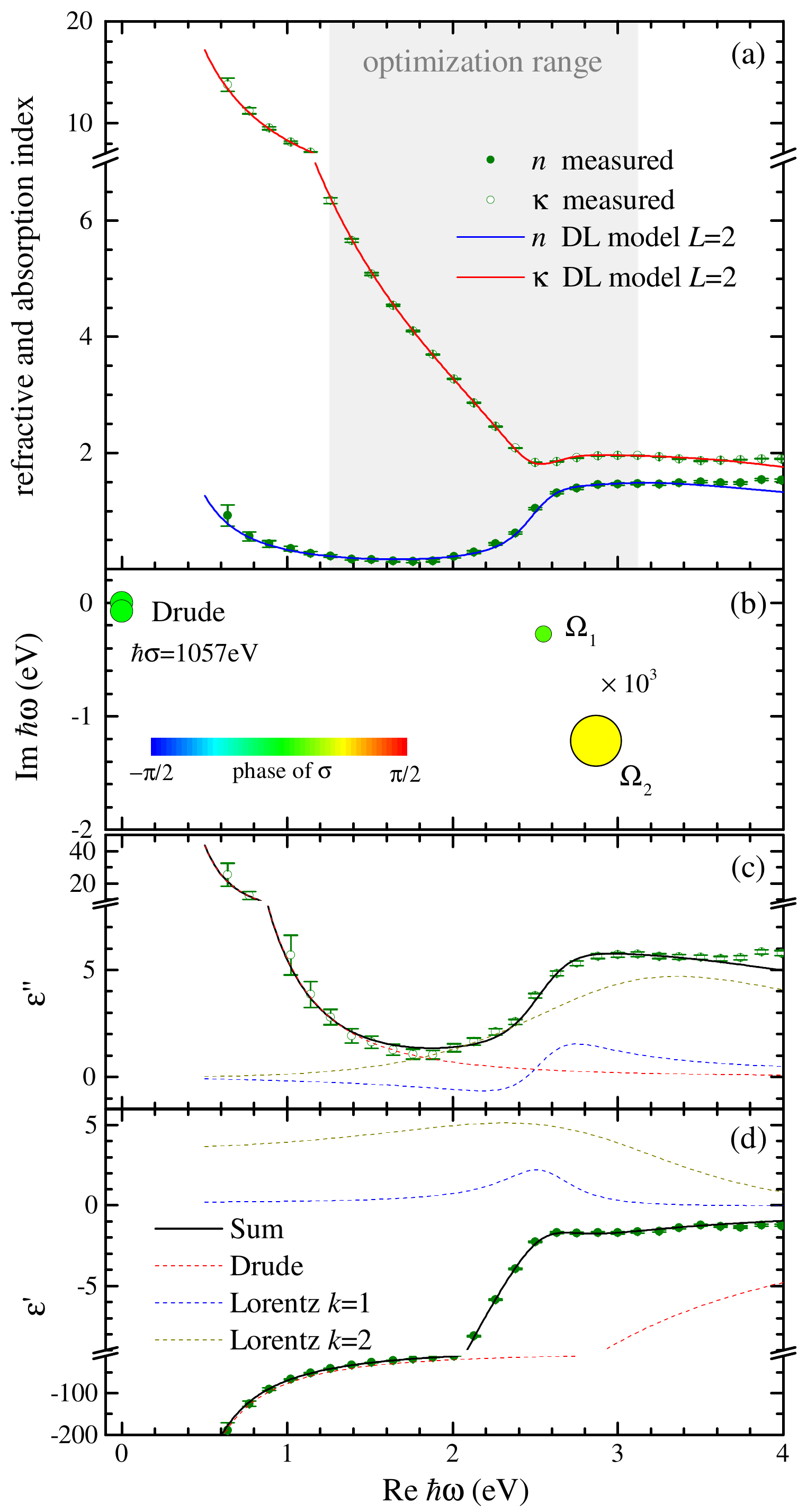}
	\caption{As \Fig{fig:JCAuL1}, but for $L=2$. Additionally, the permittivity, $\epsr$ and $\epsi$, is shown in (c) and (d), together with the individual terms of the model \Eq{eqn:eps2}.}
	\label{fig:JCAuL2}
\end{figure}

The parameter space volume to be covered in such a procedure increases exponentially with $L$, making it computationally prohibitive to use this approach for large $L$. Increasing $L$, we therefore revert to a different strategy. Instead of guessing all $\Omega_k$ randomly, we use the optimized values for  $\Omega_1,...,\Omega_{L-1}$ of the model with $L-1$ poles as starting values for the simulation for $L$ poles, and choose the starting value for the additional pole as $\Omega_L=[1-{i}/{(L+1)}]\omega_N$. This procedure is sketched in \Fig{fig:Guess}. It is fast but can result in not finding the global minimum. However, we can vary the range of the experimental data to be fitted in order to provide more starting points. Here, we choose to keep the lowest frequency $\omega_1$ fixed but vary $\omega_N$ and consequently $N$. Increasing or decreasing $N$ by one, we use as starting point the optimized values for $N$.

Furthermore, going back, from $L+1$ to $L$, just removing one pair of Lorentz poles provides $L+1$ additional starting values for the simulation with $L$ poles. It is also possible to go back multiple steps, e.g., from $L+2$ to $L$ provides $(L+2)(L+1)/2$ starting values -- this however has not been used to produce the $S$ values in this paper.

Remaining abrupt changes of $S$ with $N$ can (but do not have to) indicate that the global minimum was not yet obtained, and more starting values should be employed.

\section{Results for Gold}
\label{sec:Results}

Here we discuss examples of the DL model optimized for measured material dispersions. As the main example we use the data for gold by Johnson and Christy\Cite{JohnsonPRB72}, which is widely used in the literature and we can compare our model with previous approaches. Fits for other materials presented in Ref. \Cite{JohnsonPRB72}, such as silver and copper, as well as for a newer measurement on single-crystalline gold\Cite{BabarAO15}, are provided in the Appendix.

The data by Johnson and Christy\Cite{JohnsonPRB72} covers the $\hbar\omega$ range from  0.64\,eV to 6.6\,eV, and provides $n$ and $\kappa$ with their errors as discussed in \Sec{sec:ExpData}. Previous works\Cite{VialAPA11,VialJPD07} concentrated on a narrower region 1.24\,--\,3.1\,eV, corresponding to the extended visible light range 400\,--\,1000\,nm. We start by using this range for the optimization, as it is the most relevant range for applications, and also allows us to directly compare our results with previous findings.
We use $D=1$, which is sufficient in the frequency range considered, as the photon frequency is much higher than the Drude damping, i.e., $\omega\gg\gamma$.

The optimized model using $L=1$ is compared with the experimental data in \Fig{fig:JCAuL1}(a).
The refraction and absorption indices are shown as functions of the photon energy, with the measured data including error bars and lines representing the fit functions of the DL model. The poles of the model [see Eqs.(\ref{Drude}) and (\ref{eqn:eps2})] are shown as circles in \Fig{fig:JCAuL1}(b), centered at their pole positions $\Omega_j$ in the complex photon energy plane, with the complex pole weight represented by the circle area proportional to $|\sigma_j|$ and the color giving the phase. We find $S=2.47$ for this fit, with other parameters given in \Tab{tab:para}. One can see the dominant contribution of the Drude pole having the weight about 200 times larger than the Lorentz pole. We can also see that the Lorentz pole is properly positioned to model the interband transitions of gold above 2.3\,eV. The phase of $\sigma_1$ is $\pi/4$ close to the phase $\pi/2$ of a classical damped Lorentz oscillator, such as a mass on a spring. The resonance is at $\Omega_1' \sim 2.7$\,eV, around the center of the interband transition within the optimization range, and the half-width of the resonance, $-\Omega_1'' \sim 0.7$\,eV, is approximately covering the width of the interband transitions in the same range.
Comparing the model with the data in \Fig{fig:JCAuL1} we can see that using only a single Lorentz pole is insufficient to describe the measured data within their error, which is confirmed by the corresponding value of $S$ above unity.

\begin{table}[t]
  \begin{tabular}{>{$}c<{$} | >{$}c<{$} | >{$}c<{$} | >{$}c<{$}| >{$}c<{$}}
	  L   			& 1	           & 2		& 3			& 3\\
    \hline
 \rule{0pt}{3ex}
     \varepsilon_\infty	& 3.9199	& 2.6585	& -10.534		& 1\\
      \gamma ({\rm eV})	& 0.0893	& 0.07247	& 0.07373		& 0.074018\\
      \sigma ({\rm eV})	& 875.79	& 1056.9	& 997.41		& 995.13\\
   \hline
  \rule{0pt}{3ex}
    \Omega_1' ({\rm eV})	& 2.7326	& 2.5509	& 2.5997		& 2.6039\\
      \Omega_1'' ({\rm eV})	& -0.69021	& -0.27427	& -0.43057		& -0.42417\\
      \sigma_1' ({\rm eV})	& 3.0701	& 0.57604	& 1.4835		& 1.4145\\
      \sigma_1'' ({\rm eV})	& 2.9306	& 0.18443	& 0.88382		& 0.89754\\
   \hline
\rule{0pt}{3ex}
      \Omega_2' ({\rm eV})	& -  		& 2.8685	& 3.7429		& 3.685\\
      \Omega_2'' ({\rm eV})	& -		& -1.2195	& -1.2267		& -1.2475\\
      \sigma_2' ({\rm eV})	& -		& 4.1891	& 1.1372		& 1.5109\\
      \sigma_2'' ({\rm eV})   & -		& 4.2426	& 3.8223		& 3.9555\\
   \hline
 \rule{0pt}{3ex}
     \Omega_3' ({\rm eV})	& -		& -		& 7.3145		& 17.087\\
      \Omega_3'' ({\rm eV})	& -		& -		& -21.843		& -0.41705\\
      \sigma_3' ({\rm eV})	& -		& -		& 225.27		& -30.678\\
      \sigma_3'' ({\rm eV})	& -		& -		& -193.27		& 13.987\\
  \hline
 \rule{0pt}{3ex}
     \hbar\omega_1 ({\rm eV})	& 1.24	& 1.24	&0.64		& 0.64\\
      \hbar\omega_N ({\rm eV})	& 3.1		& 3.1		&6.6		& 6.6\\
      2N					& 30		& 30		&98		& 98\\
      \text{Fit parameters}		& 7		& 11		&15		& 14\\
      S					& 2.4735	& 1.0029	&1.4747	& 1.4872\\
  \end{tabular}
  \caption{Optimized model parameters for different number of Lorentz pole pairs $L$ and optimization energy ranges corresponding to the data shown in Figs.\,\ref{fig:JCAuL1}, \ref{fig:JCAuL2}, and \ref{fig:JCAuL3}. The number of data values $2N$, the number of fit parameters, and the resulting error $S$ are also given. The last column shows an example where we choose $\varepsilon_\infty=1$.}
  \label{tab:para}
\end{table}

\begin{figure}[t]
	\includegraphics[width=\columnwidth]{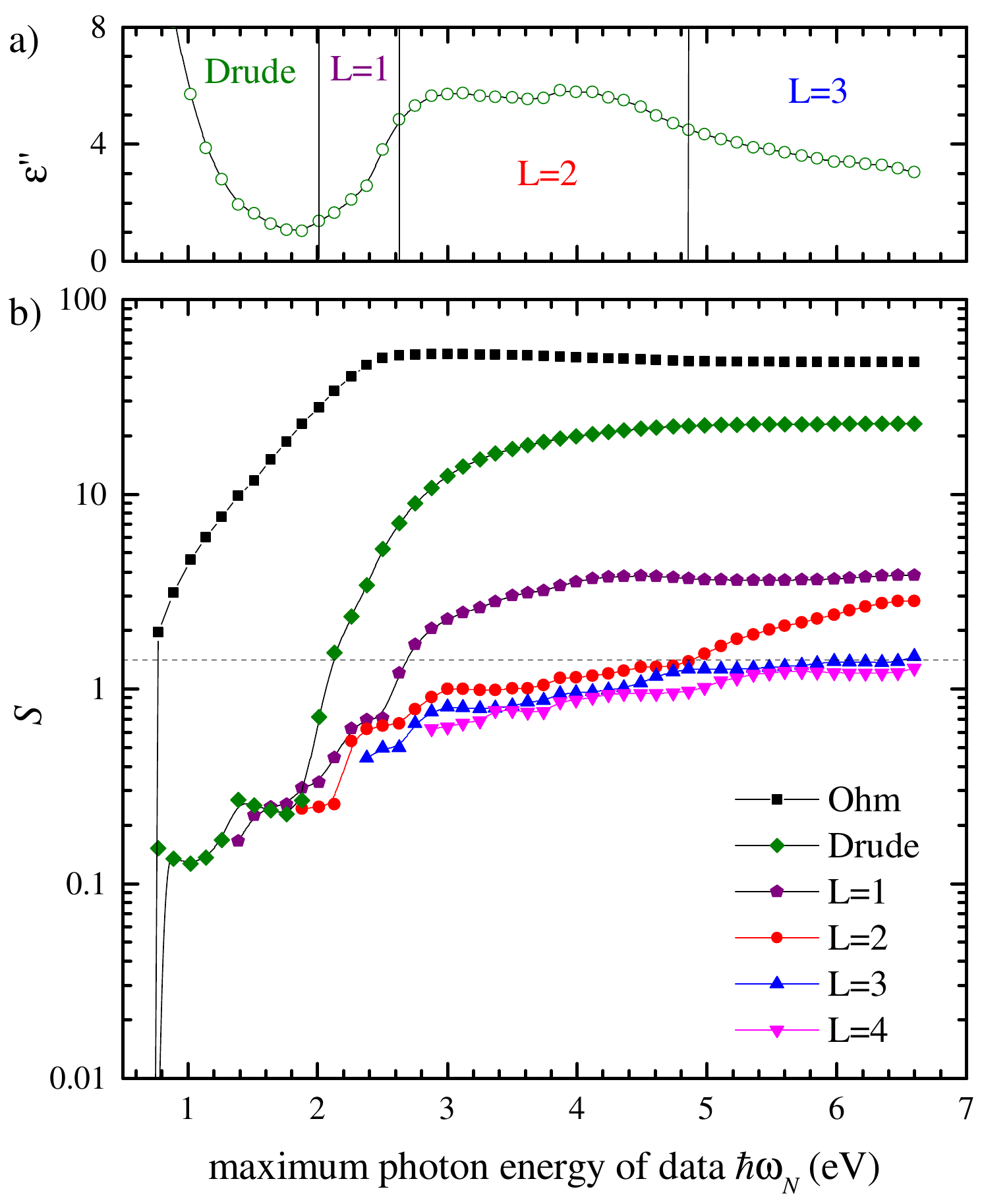}
	\caption{(a) $\mepsi$ as function of $\hbar\omega_j$. (b) Error $S$ as function of the upper photon energy limit of the fitted data range for Au data taken from \Onlinecite{JohnsonPRB72}. Results for various number of poles in the model are given. Lines are guides for the eye. The maximum photon energy ranges suited for the different  number of poles are indicated in (a) by vertical lines.}
	\label{fig:JCAuS}
\end{figure}

Moving to a model with two pairs of Lorentz poles, $L=2$, the error is decreased to $S=1.0$. The value of $S$ below $\sqrt{2}$ indicates that this is sufficient to model the data in the optimization range. This is also seen in \Fig{fig:JCAuL2}, with the corresponding parameters given in \Tab{tab:para}. We show in Figs.\,\ref{fig:JCAuL2}(c) and (d) the data and the fit also for $\epsr$ and $\epsi$, the quantities which are actually fitted, according to \Eq{eqn:Error}. Individual pole contributions to $\epsr$ and $\epsi$ are displayed as well.
The interband transitions are now described by two Lorentz poles. The first pole is at $\Omega_1'\sim 2.6$\,eV, close to the onset of the interband transition region, with a half-width of $-\Omega_1'' \sim 0.3$\,eV. This pole describes the edge of the interband transitions. Indeed, it has a phase close to zero, which is appropriate to describe the asymmetry of the edge, see \Fig{fig:JCAuL2}(c).  The second pole is at $\Omega_2'\sim 2.9$\,eV, with a half-width of $-\Omega_2'' \sim 1.2$\,eV. This pole describes the central part of the interband transition region. It has a weight about ten times higher than the first pole, and a phase close to $\pi/4$. The contribution to $\epsi$ has accordingly a peak at around the resonance, while the contribution to $\epsr$ is more dispersive.

Concerning the relation of the poles to intraband transitions in solids, it is important to emphasize that in microscopic theory the optical response is due to a large number of transitions, often described by a continuum. This continuum, however, can be represented by an infinite or a finite number of poles of the self-energy describing the effects of screening and frequency dispersion. Therefore, the model with a limited number of Lorentz oscillators presents a fully physical though approximate approach, collecting the oscillator strength and transition energies of the continuum into a finite number of poles. The resulting pole positions and weights depend on the energy range to be described and represent sets of microscopic transitions in the material.

As we have seen, we can optimize the model parameters for a given photon energy range and quantify the fit quality by the resulting value of $S$.
Now we use a variable optimization range, from the lowest measured photon energy to a variable upper boundary of the photon energy $\hbar\omega_N$ taking all available measured values. We show the resulting $S$ values in \Fig{fig:JCAuS}(b) for different numbers of poles taken into account. We can see that with an increasing number of poles, the error $S$ is decreasing, as expected considering the increasing number of parameters. Instead, increasing $\hbar\omega_N$ results in larger values of $S$, since a model of a given number of parameters is used to describe an increasing number of data.

When keeping only the $\omega=0$ pole, corresponding to an Ohm's law dispersion, the error is always above $\sqrt{2}$. This is expected, as Ohm's law is suited only to describe the dispersion at photon energies well below the Drude damping, which is about 0.1\,eV for gold. Moving to two poles, representing the Drude model, we see that the error stays below unity until $\hbar\omega_N$ approaches the interband transitions, seen in \Fig{fig:JCAuS}(a) as a region of increasing $\epsi$ above 2\,eV. This shows that the Drude model is well suited to describe the measured data, as long as the influence of the higher energy electronic transitions can be represented simply by a background constant $\varepsilon_\infty$.
Adding the first pair of Lorentz poles ($L=1$), the effect of the interband transitions can be described up to about 2.6\,eV, where the plateau in $\epsi$ commences. Adding the second pair of Lorentz poles ($L=2$), the effect of the interband transitions can be described up to about 4.9\,eV, where $\epsi$ starts to decrease. Adding the third pair of Lorentz poles ($L=3$) allows us to adequately describe the full range of measured data up to 6.5\,eV.

\begin{figure}[t]
\includegraphics[width=\columnwidth]{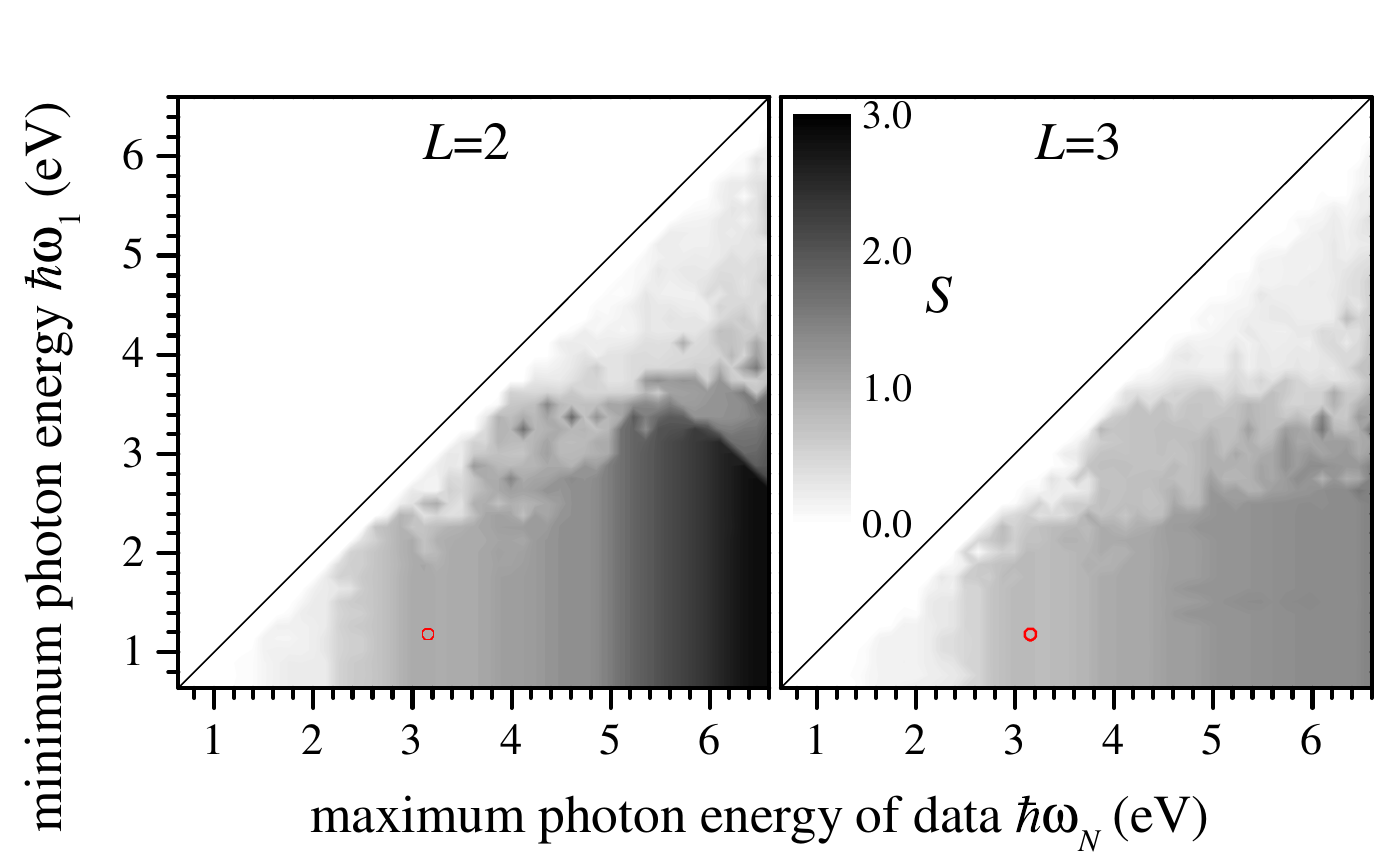}
\caption{Error $S$ for $L=2$ and $L=3$ on a gray scale as given as a function of both the lower and upper photon energy limits of the fitted data range for the Au data in \Onlinecite{JohnsonPRB72}. The circle indicates the range $1.24\leqslant \hbar\omega \leqslant3.1$\,eV used in \Onlinecite{VialJPD07}. }
\label{fig:JCAuS2D}
\end{figure}

Figure~\ref{fig:JCAuS2D} shows the error $S$ for $L=2$ and $L=3$, and both lower and higher limits of the fitted range changing. Including more data points results in higher errors, as seen by the gradient of $S$ towards the lower right corner. We can see that any range of the available data can be described by the DL model with $L=3$ with errors $S<\sqrt{2}$.
The region of interest used in previous works\Cite{VialAPA11,VialJPD07}, $1.24\leqslant \hbar\omega \leqslant3.1$\,eV, is also indicated by red circles.
Using the parameters of \Onlinecite{VialJPD07}, corresponding to the model with $L=2$, we find $S=1.96$, which is larger than the value $S=1$ we found (see \Tab{tab:para}). This can be attributed to the fact that in \Onlinecite{VialJPD07} the absolute error of $\eps$ was minimized, not taking into account the experimental errors. Such a minimization corresponds in our case to setting $\epsrt=\epsit=1$ for all data points. Using these errors, both in the definition of $S$ and in the optimization of the parameters, we find $S=0.019$ for $L=2$, while using the parameters of \Onlinecite{VialJPD07} results in $S=0.028$. This confirms the high quality of our optimization method.

\begin{figure}[t]
	\includegraphics[width=\columnwidth]{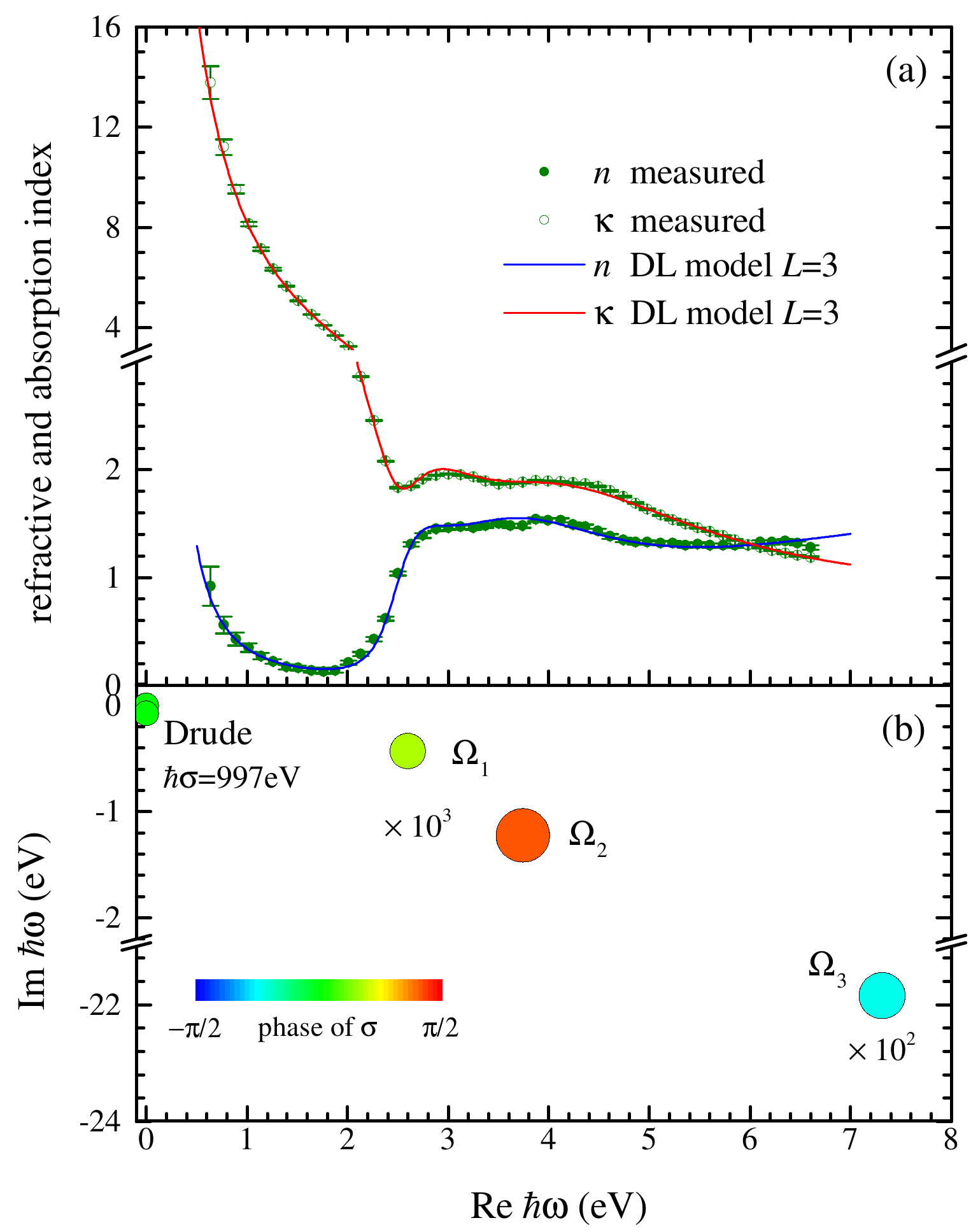}
	\caption{As \Fig{fig:JCAuL1}, but for $L=3$ and optimized for the full data range of $\hbar\omega$ given in \Onlinecite{JohnsonPRB72}, from 0.64\,eV to 6.6\,eV.}
	\label{fig:JCAuL3}
\end{figure}

Finally, the model with $L=3$ optimized for the full data range is compared with the measured data in \Fig{fig:JCAuL3}. The fitted parameters are given in \Tab{tab:para}. We see that the first two Lorentz poles are similar to those in the $L=2$ model used for the limited range and shown in \Fig{fig:JCAuL2}. To describe the full range, an additional pole  at higher energy, having $\Omega_3'\sim 7.3$\,eV and a half-width of $-\Omega_3'' \sim 21.8$\,eV, is needed. This pole describes the continuum of interband transitions, and takes over the role of $\varepsilon_\infty$, which in this fit has a value below $\sqrt{2}$. The weight of the pole is about ten times higher than for the second pole, and the phase is close to $-\pi/4$. Fixing $\varepsilon_\infty=1$, which is well suited for FDTD methods, the main difference (see \Tab{tab:para}) is a change in the high energy third pole.

\begin{figure}[t]
\includegraphics[width=\columnwidth]{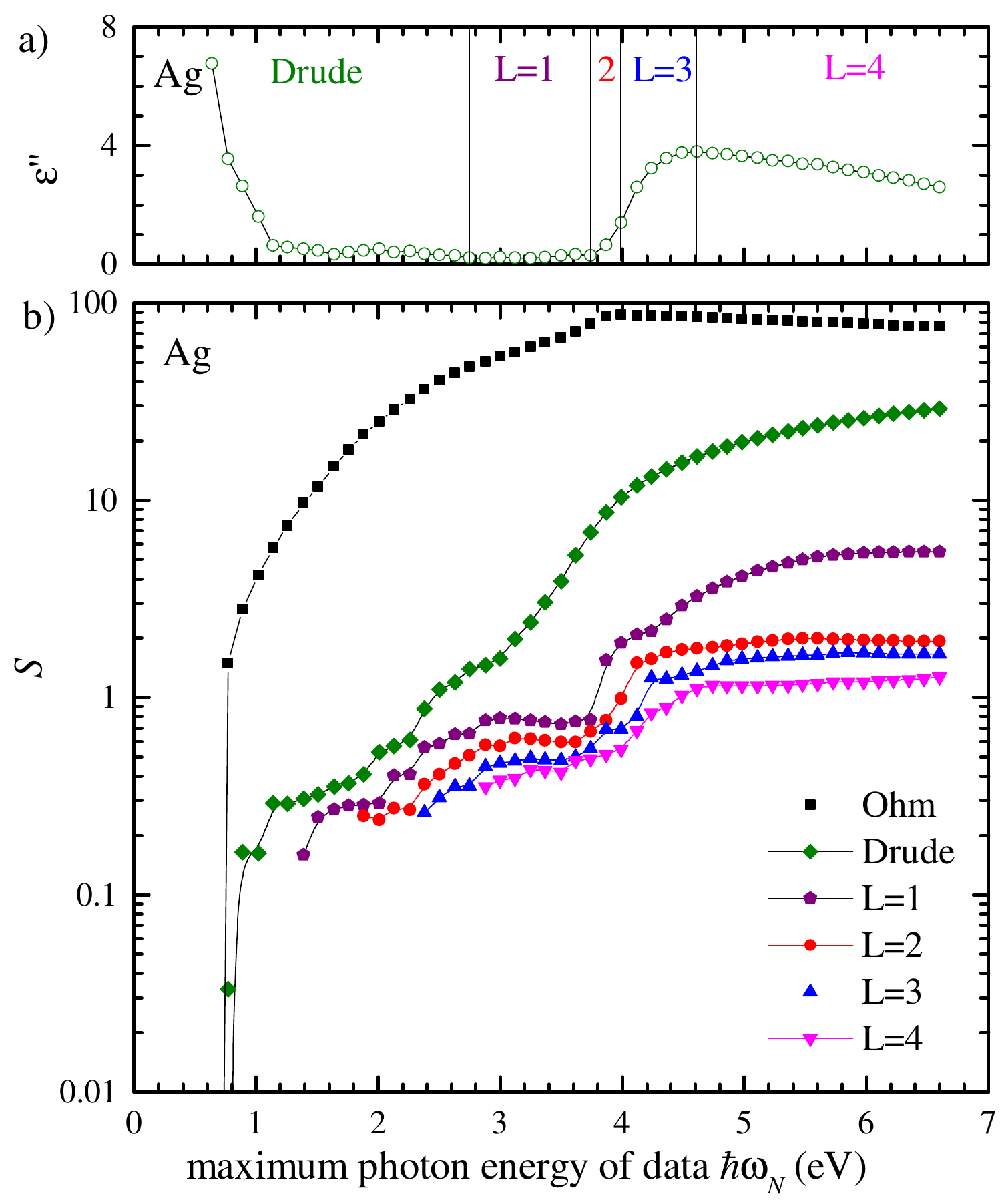}
\caption{As \Fig{fig:JCAuS}, but for silver.}
\label{fig:JCAgS}
\end{figure}

\section{Conclusion}

In conclusion, we have presented an optimization algorithm to determine the parameters of a generalized Drude-Lorentz model for the permittivity of materials. For $L$ pairs of Lorentz poles and $D$ Drude poles taken into account, the developed algorithm uses an analytic minimization over the $2L+D+1$ linear parameters of the model (the generalized conductivities and high frequency value $\varepsilon_\infty$), and a gradient decent method for determining the $2L+D$ nonlinear parameters of the model (the Drude and Lorentz pole frequencies), with a suited choice of the starting values, resulting in fast and reliable determination of the best global fit.

Comparing our results with previous literature\Cite{VialJPD07}, we find that the weighted error is improved by a factor of two for the same number of poles. For gold, we find that the Drude model is sufficient up to photon energies of 2\,eV, one additional pair of Lorentz oscillators up to 2.6\,eV, two up to 4.8\,eV, and three up to 6.5\,eV. We provide parameters for more materials in the Appendix, including a recent dataset for monocrystalline gold\Cite{BabarAO15}.

The optimization program implementing the described algorithm is provided\Cite{Program} for modelling any measured data for the refractive index or permittivity. The data presented in this work are available from the Cardiff University data archive\Cite{paperCUdata}.

\acknowledgments 

This work was supported by the S\^er Cymru National Research Network in Advanced Engineering and Materials. E.A.M. acknowledges support by RBRF (Grant No. 16-29-03282).

\appendix

\section{Results for silver, copper, and single-crystalline gold}
\label{A1}

\begin{figure}
	\includegraphics[width=\columnwidth]{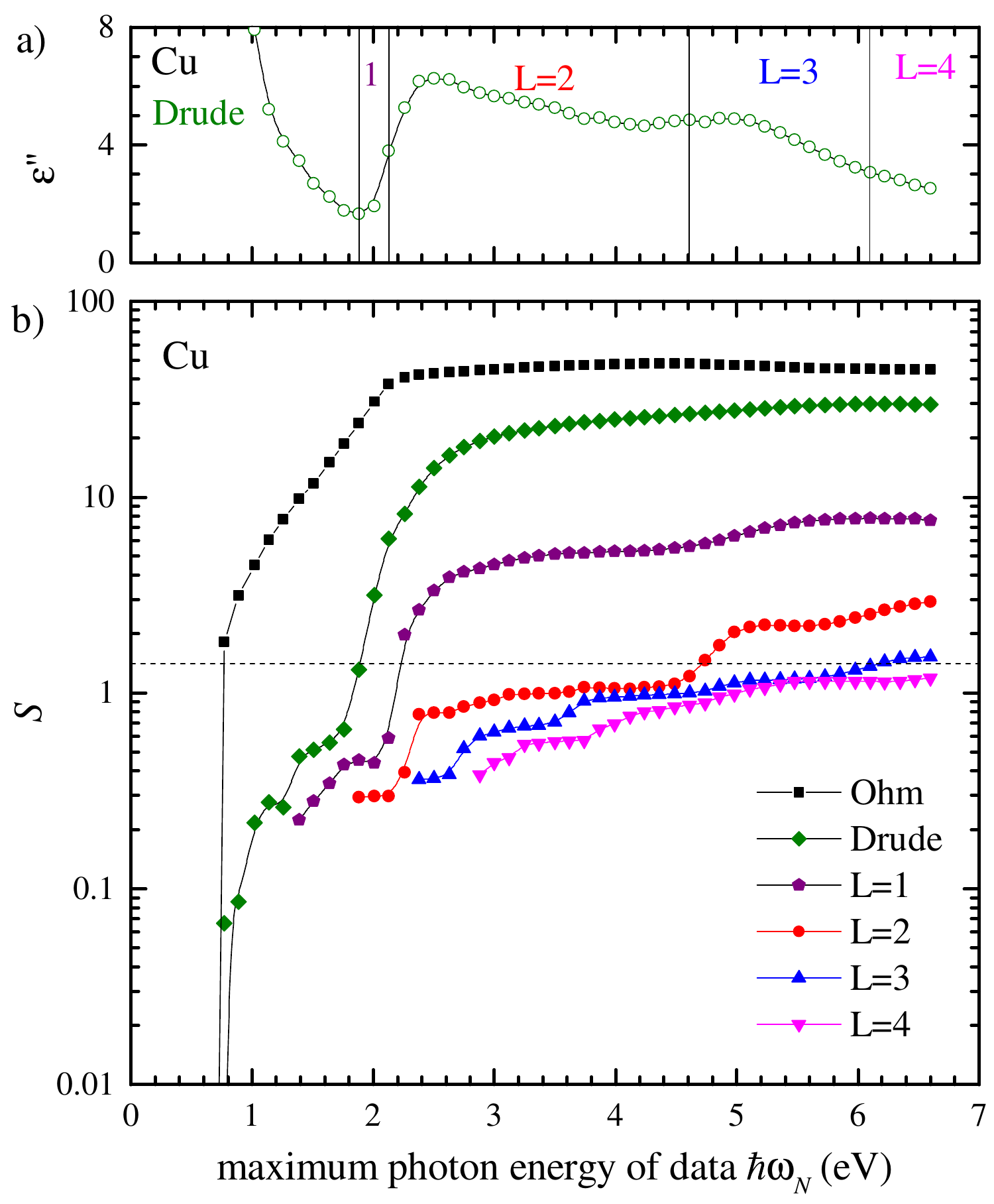}
	\caption{As \Fig{fig:JCAuS}, but for copper.}
	\label{fig:JCCuS}
\end{figure}

\begin{figure}
	\includegraphics[width=\columnwidth]{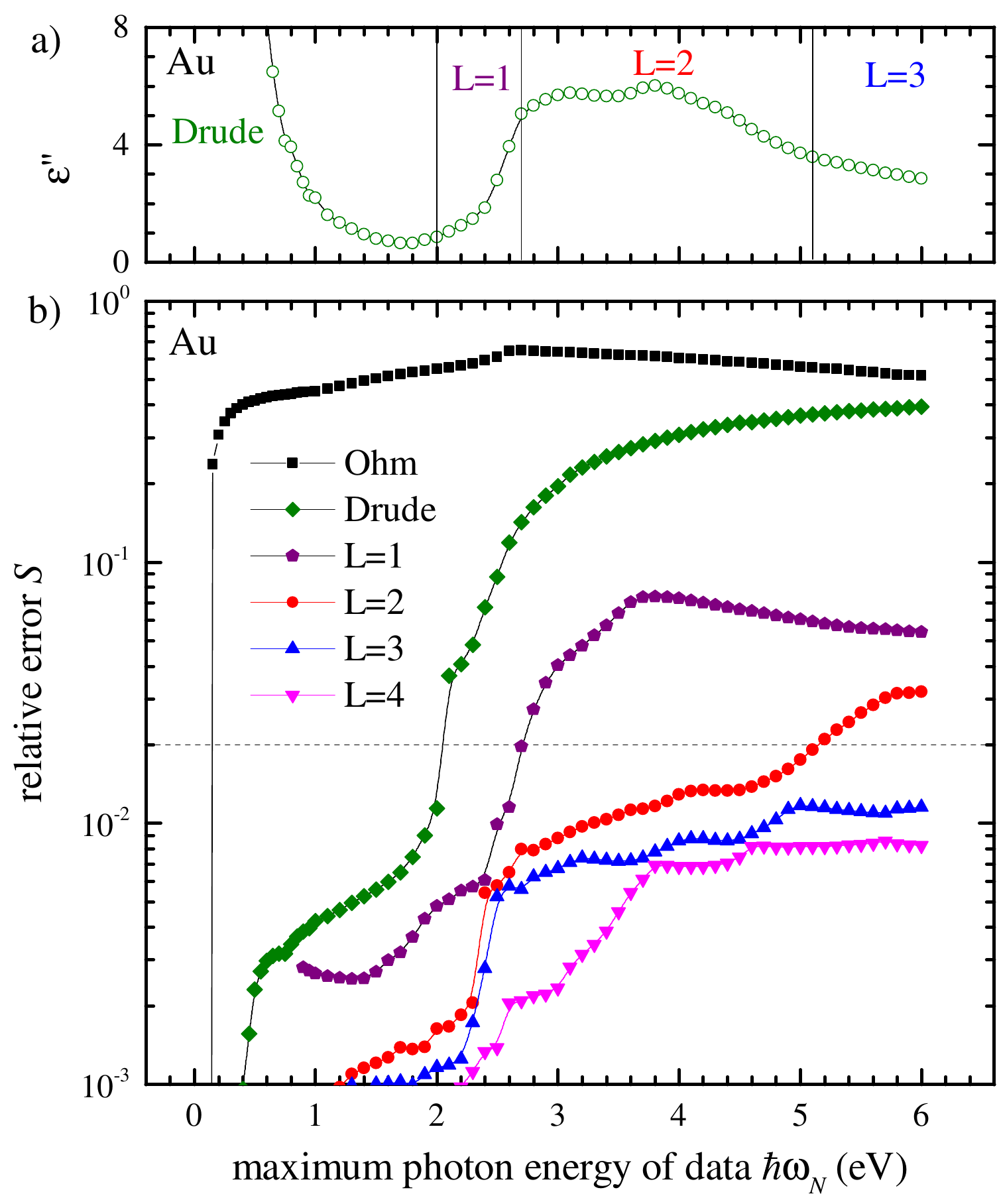}
	\caption{As \Fig{fig:JCAuS}, but for gold using data from \Onlinecite{BabarAO15}. We show a dashed line at 2\% relative error as a guide to a satisfactory fit.  }
	\label{fig:BWAuS}
\end{figure}

\begin{table}
  \begin{tabular}{>{$}c<{$} | >{$}c<{$} | >{$}c<{$} | >{$}c<{$} | >{$}c<{$}}
	  \text{Material}				& \text{Ag}       & \text{Cu}		& \text{Au}       & \text{Au}\\
	  L   					& 4         & 4			& 3        & 4\\
    \hline
 \rule{0pt}{3ex}
     \varepsilon_\infty	& 0.77259	&12.294	& 1.1584      &0.83409\\
      \gamma ({\rm eV})	& 0.02228	& 0.07044	& 0.02321	& 0.02334\\
      \sigma ({\rm eV})	& 3751.4	& 1137.9	& 3155.3	& 3134.5\\
    \hline
\rule{0pt}{3ex}
      \Omega_1' ({\rm eV})	&3.9173	& 2.1508	&2.1339	& 2.6905\\
      \Omega_1'' ({\rm eV})	&-0.06084	&-0.23449	&-3.4028	& -0.16645\\
      \sigma_1' ({\rm eV})	&0.09267	& 0.95283	&12.0		& -0.01743\\
      \sigma_1'' ({\rm eV})	& 0.01042	& -0.12983	&-5.5574	&0.3059\\
    \hline
\rule{0pt}{3ex}
      \Omega_2' ({\rm eV})	& 3.988	& 4.6366	& 2.6319	& 2.8772\\
      \Omega_2'' ({\rm eV})	& -0.04605	&-0.68811	& -0.33701	& -0.44473\\
      \sigma_2' ({\rm eV})	& -0.0015342	& 0.97953	&1.0547	& 1.0349\\
      \sigma_2'' ({\rm eV})   & -0.062233	& 0.48395	&0.53584	& 1.2919\\
    \hline
\rule{0pt}{3ex}
      \Omega_3' ({\rm eV})	& 4.0746	&4.9297	&4.0803	& 3.7911\\
      \Omega_3'' ({\rm eV})	& -0.63141	& -4.6932	&-0.99872	& -0.81981\\
      \sigma_3' ({\rm eV})	& 1.4911	& -61.583	&-1.3103	& 1.2274\\
      \sigma_3'' ({\rm eV})	& 0.40655	& 35.021	&2.7819	& 2.5605\\
    \hline
\rule{0pt}{3ex}
      \Omega_4' ({\rm eV})	& 4.6198	& 8.8317	& -		& 4.8532\\
      \Omega_4'' ({\rm eV})	& -2.8279	& -0.2679	& -		& -13.891\\
      \sigma_4' ({\rm eV})	& 4.2843	& -12.186	& -		& 9.85\\
      \sigma_4'' ({\rm eV})	& 4.2181	& 5.1474	& -		& 37.614\\
    \hline
\rule{0pt}{3ex}
     \hbar\omega_1 ({\rm eV})	& 0.64	& 0.64	&0.1	           & 0.1\\
      \hbar\omega_N ({\rm eV})	& 6.6		& 6.6		&6.0	           & 6.0\\
      N					& 49		& 49		&69		& 69\\
      S					& 1.2684	& 1.0956	&0.01151       & 0.00826\\
  \end{tabular}
  \caption{
  As Table~\ref{tab:para}, but for the data\Cite{JohnsonPRB72} for Ag and Cu and for the data\Cite{BabarAO15} for Au, corresponding to the full fit range shown in Figs.\,\ref{fig:JCAgS}, \ref{fig:JCCuS}, and \ref{fig:BWAuS}, respectively.
  }
  \label{tab:para2}
\end{table}

Here we show the results of the fit of the DL model for other materials. As in \Fig{fig:JCAuS} of the main text, we use a variable upper limit $\hbar\omega_N$ of the optimization range and show the resulting $S$ values for different numbers of poles. In \Fig{fig:JCAgS} we show results for silver using the data from \Onlinecite{JohnsonPRB72}, having the lower photon energy limit at $\hbar\omega_1=0.64$\,eV.
Ag has interband transitions above 4\,eV.
We find that the Drude model is sufficient up to photon energies of 2.4\,eV, one additional pair of Lorentz oscillators up to 3.7\,eV, two up to 4.0\,eV, three up to 4.7\,eV, and four up to a value above the upper limit of 6.6\,eV.

In \Fig{fig:JCCuS} we show results for copper using the data from \Onlinecite{JohnsonPRB72}, having the lower photon energy limit at $\hbar\omega_1=0.64$\,eV. Cu has interband transitions above  2\,eV.  We find that the Drude model is sufficient up to photon energies of 1.9\,eV, one additional pair of Lorentz oscillators up to 2.2\,eV, two up to 4.7\,eV and three up to 6\,eV.

In \Fig{fig:BWAuS} we show results for gold using the newer experimental data from \Onlinecite{BabarAO15}. This data does not provide the experimental error. We therefore have chosen here to minimize the relative error instead, using $\epst=\eps$ in \Eq{eqn:Error}.
We see a similar behavior as for the data from \Onlinecite{JohnsonPRB72}, see \Fig{fig:JCAuS}.
The parameters fitted for the full spectral range shown in Figs.\,\ref{fig:JCAgS}, \ref{fig:JCCuS}, and  \ref{fig:BWAuS}, with $L=3$ and 4, are given in \Tab{tab:para2}.
The model parameters for the data presented in Figs.\,\ref{fig:JCAgS}, \ref{fig:JCCuS}, and \ref{fig:BWAuS} are available in \Onlinecite{paperCUdata}.


\begin{thebibliography}{10}

\bibitem{VialJPD07}
A. Vial and T. Laroche, J. Phys. D: Appl. Phys. {\bf 40},  7152  (2007).

\bibitem{MuljarovPRB16}
E.~A. Muljarov and W. Langbein, Phys. Rev. B {\bf 93},  075417  (2016).

\bibitem{JohnsonPRB72}
P.~B. Johnson and R.~W. Christy, Phys. Rev. B {\bf 6},  4370  (1972).

\bibitem{RakicAO98}
A.~D. Raki\'{c}, A.~B. Djuri$\check{\rm s}$i\'{c}, J.~M. Elazar, and M.~L.
  Majewski, Appl. Opt. {\bf 37},  5271  (1998).

\bibitem{EtchegoinJCP06}
P.~G. Etchegoin, E.~C. Le~Ru, and M. Meyer, J. Chem. Phys. {\bf 125},  164705
  (2006).

\bibitem{EtchegoinJCP07}
P.~G. Etchegoin, E.~C. Le~Ru, and M. Meyer, J. Chem. Phys. {\bf 127},  189901
  (2007).

\bibitem{VialAPB08}
A. Vial and T. Laroche, Applied Physics B {\bf 93},  139  (2008).

\bibitem{VialAPA11}
A. Vial, T. Laroche, M. Dridi, and L. Le�Cunff, Appl. Phys. A {\bf 103},  849
  (2011).

\bibitem{LandauLifshitzV8Book84}
L.~D. Landau, L.~P. Pitaevskii, and E. Lifshitz, {\em Electrodynamics of
  Continuous Media}, Vol.~8 of {\em Course of Theoretical Physics}, 2nd ed.
  (Elsevier Butterworth Heinemann, Oxford, 1984).

\bibitem{AllenPRB77}
J.~W. Allen and J.~C. Mikkelsen, Phys. Rev. B {\bf 15},  2952  (1977).

\bibitem{BabarAO15}
S. Babar and J.~H. Weaver, Appl. Opt. {\bf 54},  477  (2015).

\bibitem{paperCUdata}
Cardiff University data archive: http://dx.doi.org/10.17035/d.2017.0031193886.

\bibitem{Program}
The optimization program is available at http://langsrv.astro.cf.ac.uk/DrudeLorentzFit/

\end{thebibliography}

\end{document}